\documentclass[twocolumn,amsmath,amssymb]{revtex4}
\usepackage{graphicx}
\usepackage{subfigure}
\usepackage{dcolumn}
\usepackage{bm}
\usepackage{color}
\usepackage{pstricks}
\usepackage{pst-text}
\begin{document}
\title{Exact Site and Bond Percolation Probability on Lattice-like Graphs}
\author{Marko Puljic, neuropercolation@yahoo.com}
\maketitle

\section{Exact Site Percolation on \(\mathbb{Z}^{d}\)}
\begin{figure}
\definecolor{gry}{gray}{.9}
\begin{pspicture}(4,4.2)(0,0)

\psline[linewidth=6pt,linestyle=solid,linecolor=gry]{-}(-0.2,4)(4.2,4)
\psline[linewidth=6pt,linestyle=solid,linecolor=gry]{-}(0.3,3.5)(3.7,3.5)
\psline[linewidth=6pt,linestyle=solid,linecolor=gry]{-}(0.8,3)(3.2,3)
\psline[linewidth=6pt,linestyle=solid,linecolor=gry]{-}(1.3,2.5)(2.7,2.5)

\psline[linewidth=0.5pt,linestyle=solid]{-}(0,3)(1,4)
\psline[linewidth=0.5pt,linestyle=solid]{-}(0,2)(2,4)
\psline[linewidth=0.5pt,linestyle=solid]{-}(0,1)(3,4)
\psline[linewidth=0.5pt,linestyle=solid]{-}(0,0)(4,4)
\psline[linewidth=0.5pt,linestyle=solid]{-}(1,0)(4,3)
\psline[linewidth=0.5pt,linestyle=solid]{-}(2,0)(4,2)
\psline[linewidth=0.5pt,linestyle=solid]{-}(3,0)(4,1)

\psline[linewidth=0.5pt,linestyle=solid]{-}(0,1)(1,0)
\psline[linewidth=0.5pt,linestyle=solid]{-}(0,2)(2,0)
\psline[linewidth=0.5pt,linestyle=solid]{-}(0,3)(3,0)
\psline[linewidth=0.5pt,linestyle=solid]{-}(0,4)(4,0)
\psline[linewidth=0.5pt,linestyle=solid]{-}(1,4)(4,1)
\psline[linewidth=0.5pt,linestyle=solid]{-}(2,4)(4,2)
\psline[linewidth=0.5pt,linestyle=solid]{-}(3,4)(4,3)

{\color{white}
\multiput(0,0)(2,0){2}{\circle*{0.15}} \put(3,0){\circle*{0.15}}
\put(2.5,0.5){\circle*{0.15}}
\multiput(1,1)(2,0){2}{\circle*{0.15}}
\multiput(1.5,1.5)(2,0){2}{\circle*{0.15}}
\multiput(0,2)(1,0){2}{\circle*{0.15}}
\multiput(2.5,2.5)(1,0){2}{\circle*{0.15}}
\multiput(1,3)(3,0){2}{\circle*{0.15}}
\put(2.5,3.5){\circle*{0.15}}
\multiput(0,4)(2,0){3}{\circle*{0.15}}
}
\multiput(0,0)(2,0){2}{\circle{0.15}} \put(3,0){\circle{0.15}}
\put(2.5,0.5){\circle{0.15}}
\multiput(1,1)(2,0){2}{\circle{0.15}}
\multiput(1.5,1.5)(2,0){2}{\circle{0.15}}
\multiput(0,2)(1,0){2}{\circle{0.15}}
\multiput(2.5,2.5)(1,0){2}{\circle{0.15}}
\multiput(1,3)(3,0){2}{\circle{0.15}}
\put(2.5,3.5){\circle{0.15}}
\multiput(0,4)(2,0){3}{\circle{0.15}}

\multiput(1,0)(3,0){2}{\circle*{0.15}}
\multiput(0.5,0.5)(1,0){2}{\circle*{0.15}} \put(3.5,0.5){\circle*{0.15}}
\multiput(0,1)(2,0){3}{\circle*{0.15}}
\multiput(0.5,1.5)(2,0){2}{\circle*{0.15}}
\multiput(3,2)(1,0){2}{\circle*{0.15}}
\multiput(0.5,2.5)(1,0){2}{\circle*{0.15}}
\multiput(0,3)(3,0){2}{\circle*{0.15}} \put(2,3){\circle*{0.15}}
\multiput(0.5,3.5)(1,0){2}{\circle*{0.15}} \put(3.5,3.5){\circle*{0.15}}
\multiput(1,4)(2,0){2}{\circle*{0.15}}

\put(2,2){\circle*{0.2}}

\psline[linewidth=0.5pt,linestyle=dotted,linearc=0.45]{-}(0.2,3.8)(3.8,3.8)(3.8,0.2)(3.8,0.2)(0.2,0.2)(0.2,3.8)

{\footnotesize
\put(1.82,2.2){\(\mathbf v_{0}\)}
\put(2.25,2.8){\(\mathbf 1\!\!\uparrow_{1}\)}
\put(2.75,3.3){\(\mathbf 2\!\!\uparrow_{1}\)}
\put(3.25,3.8){\(\mathbf 3\!\!\uparrow_{1}\)}
\put(2.25,1.8){\(\mathbf 1\!\!\downarrow_{2}\)}
\put(2.75,1.3){\(\mathbf 2\!\!\downarrow_{2}\)}
\put(3.25,0.8){\(\mathbf 3\!\!\downarrow_{2}\)}
\put(1.25,2.8){\(\mathbf 1\!\!\uparrow_{2}\)}
\put(0.75,3.3){\(\mathbf 2\!\!\uparrow_{2}\)}
\put(0.25,3.8){\(\mathbf 3\!\!\uparrow_{2}\)}
\put(1.25,1.8){\(\mathbf 1\!\!\downarrow_{1}\)}
\put(0.75,1.3){\(\mathbf 2\!\!\downarrow_{1}\)}
\put(0.25,0.8){\(\mathbf 3\!\!\downarrow_{1}\)}
\put(1.45,4.12){\(\mathbf 2\!\uparrow_{1}\!\!+\mathbf 2\!\uparrow_{2}\)}
}
\put(2.65,2.4){\(\mathcal A_{1}\)} \put(3.15,2.9){\(\mathcal A_{2}\)} \put(3.65,3.4){\(\mathcal A_{3}\)} \put(4.15,3.9){\(\mathcal A_{4}\)}

\end{pspicture}
\definecolor{gry}{gray}{.95}
\definecolor{gry2}{gray}{.75}
\begin{pspicture}(4.2,4.3)(0,0)
\pswedge[fillstyle=solid,fillcolor=gry,linewidth=0.5pt](2,2){2.1}{40}{140}
\pswedge[linewidth=0.5pt](2,2){2.1}{220}{320}
\psarc[linewidth=0.1pt]{-}(2,2){1.5}{40}{140}
\psarc[linewidth=0.1pt]{-}(2,2){0.7}{40}{140}

\put(2.1,1.9){\(\mathbf v_{0}\)}
\rput{24}(0.85,4){\(\mathcal A_{k}(\mathbb{Z}^{d})\)}
\rput{-40}(3.65,3.2){\footnotesize\(k\)}
\rput{-40}(3.25,2.85){\footnotesize\(j\)}
\rput{-45}(2.85,2.15){\footnotesize\(j\!-\!m\)}
\rput{156}(0.85,0){\(\mathcal A_{-k}(\mathbb{Z}^{d})\)}

\psline[linewidth=0.5pt]{->}(2,2)(1.2,3.25)
\psline[linewidth=0.5pt]{->}(2,2)(2.2,3.45)
\psline[linewidth=0.5pt]{->}(0.5,1.2)(0.65,3.57)
\psline[linewidth=0.5pt]{->}(2.4,2.55)(2.62,3.97)
\psline[linewidth=0.5pt,linecolor=gry2]{->}(1.2,3.25)(0.5,1.2)
\psline[linewidth=0.5pt,linecolor=gry2]{->}(2.2,3.45)(2.4,2.55)

\psline[linewidth=0.5pt]{->}(2,2)(0.7,0.37)
\psline[linewidth=0.5pt]{->}(2,2)(2.3,0.6) 
\psline[linewidth=0.5pt]{->}(2.9,1)(2.9,0.15)
\psline[linewidth=0.5pt,linecolor=gry2]{->}(2.3,0.6)(2.9,1)


\end{pspicture}
\caption{\label{lattice2Deg5pie4}
\(\mathbf v\!\in\!\mathbb{Z}_{4}^{2}\) is 4 edges away from \(\mathbf v_{0}\): \(\sum_{i}|a_{i}(\mathbf v)|\!=\!4\) (dotted curve).
\(j\) up-steps to  \(\mathcal A_{j}\) and then \(m\) down-steps to \(\mathcal A_{j-m}\) and then \((m\!+\!k\!-\!j)\) up-steps to \(\mathcal A_{k}\) sum up to \((k\!+\!2m)\) steps (right).
}
\end{figure}
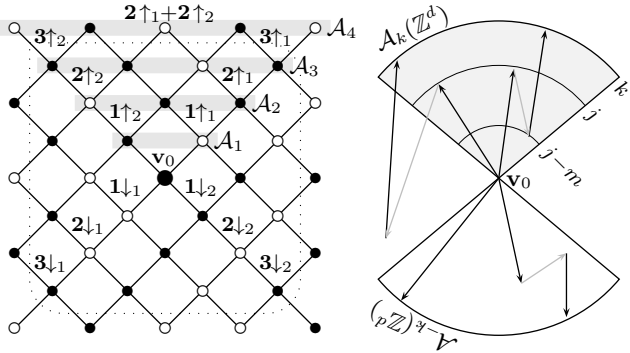

\noindent
Vertices (sites), open with the smallest probability \(p_{H}\), percolate when they form an infinite open path from graph's origin \(\mathbf v_{0}\), \footnote{Path is a walk via edges visiting each vertex only once.}.
Usually, \(p_{H}\) values are approximated, but there are a few instances of special lattices with the exact results, \citep{,hammersley1961,kesten1980,bollobas2006}.
\\ \\
In finite graph \(\mathbb{Z}_{k}^{d}\), \(d\) pairs of opposite arcs are \(k\) edges away from \(\mathbf v_{0}\), Fig. \ref{lattice2Deg5pie4} (left).
Basis \(\mathcal B\left(\mathbb{Z}^{d}\right)\) and the integers \(a_{i}(\mathbf v)\) assign the place to vertex \(\mathbf v\in\mathbb{Z}_{k}^{d}\):
\begin{align}
\mathbf v_{0}&=0=\text{vertex at origin of }\mathbb{Z}_{k}^{d}\subset\mathbb{Z}^{d} \notag \\
\mathcal B\left(\mathbb{Z}^{d}\right)&=\left\{\uparrow_{1},\uparrow_{2},\uparrow_{3},..,\uparrow_{d}\right\}\ \&\ -\!\!\uparrow_{i}=\downarrow_{i} \notag \\
\mathbf v&=a_{1}\!\uparrow_{1}+a_{2}\!\uparrow_{2}+a_{3}\!\uparrow_{3}+..+a_{d}\!\uparrow_{d} \notag \\
\lVert\mathbf v\rVert&=\displaystyle\sum_{i=1}^{d}|a_{i}(\mathbf v)|\ \&\ \lVert\mathbf v\rVert\le k\ \forall \mathbf v\!\in\!\mathbb{Z}_{k}^{d} \notag
\end{align}
\(\mathbf v\)'s neighbors can be partition into \(d\) up-step neighbors traversed via \(\uparrow_{i}\) and \(d\) down-step neighbors traversed via \(\downarrow_{i}\), so that the shortest traversal from \(\mathbf v_{0}\) to arc \(\mathcal A_{k}\left(\mathbb{Z}^{d}\right)\) is a traversal via up-step neighbors.
Arcs in \(\mathbb{Z}_{k}^{d}\) look the same, and by rotation of \(\mathbb{Z}^{d}\), any arc can be \(\mathcal A_{k}\left(\mathbb{Z}^{d}\right)\):
\begin{align}
\mathcal N_{u}\!\left(\mathbf v,\mathbb{Z}^{d}\right)&=\text{up-step neighbors of }\mathbf v\!=\!\{\mathbf v+\!\!\uparrow_{1},..,\mathbf v+\!\!\uparrow_{d}\} \notag \\
\mathcal A_{k}\left(\mathbb{Z}^{d}\right)&=\!\!\!\!\displaystyle\bigcup_{\mathbf v\in\mathcal A_{k-1}\left(\mathbb{Z}^{d}\right)}\!\!\!\!\mathcal N_{u}\left(\mathbf v,\mathbb{Z}^{d}\right):\mathcal A_{1}\left(\mathbb{Z}^{d}\right)=\mathcal B\left(\mathbb{Z}^{d}\right) \notag \\
\mathbf v&\in\mathcal A_{k}\left(\mathbb{Z}^{d}\right)\Rightarrow\lVert\mathbf v\rVert=k \notag
\end{align}

\subsection{Upper and Lower Bound for \(p_{H}\left(\mathbb{Z}^{d}\right)\)}
\noindent
If there is an open path from \(\mathbf v_{0}\) to \(\mathbf v\!\in\!\mathcal A_{k}(\mathbb{Z}^{d}):k\to\infty\), \(\mathbb{Z}^{d}\) percolates.
The shortest paths from open \(\mathbf v_{0}\) to \(\mathcal A_{k}(\mathbb{Z}^{d})\) are \(k\)-paths built by the up-step traversal.
After the first up-step to \(\mathcal A_{1}(\mathbb{Z}^{d})\), the number of paths \(d\)-tuples, so there are \(d\) 1-paths and the expected number of percolating paths in graph induced by the up-step traversal is \(\psi_{1}=dp\).
After second up-step, there are \(d^{2}\) 2-paths and \(\psi_{2}=(\psi_{1}d)\cdot p=(dp)^{2}\).
Inductively, after \(k\) up-steps, there are \(d^{k}\) \(k\)-paths and \(\psi_{k}=(dp)^{k}\):
\begin{align}
k\text{-path}&=\text{path from }\mathbf v_{0}\text{ to }\mathbf v\!:\!\mathbf v\!\in\!\mathcal A_{k}\!\left(\mathbb{Z}^{d}\right)\ \&\ \lVert\mathbf v\rVert\!=\!k  \notag \\
(k\!+\!m)\text{-path}&=k\text{-path exdended by \(m\) edges} \notag \\
n_{k}\!\Big(\!\mathcal A_{k}\!\left(\mathbb{Z}^{d}\right)\!\Big)&=\text{number of \(k\)-paths}=d^{k} \notag \\
\psi\left(\mathbb{Z}^{d},p\right)&=
\begin{array}{l}
\text{number of percolating} \\
\text{paths in }\mathbb{Z}_{k}^{d}\text{ for }k\to\infty
\end{array}
\ge\psi_{k}=(dp)^{k} \notag \\
\Rightarrow&p_{H}\left(\mathbb{Z}^{d}\right)\le\displaystyle\frac{1}{d}\label{e1}
\end{align}
If each \(k\)-path is extended by 1 down-step and 1 up-step, avoiding vertex repetition, to \((k\!+\!2)\)-path, there would be no more than \(d^{k}\) \((k\!+\!2)\)-paths, Fig. \ref{lattice2Deg5pie4} (right).
Other extensions would have to come from the non \(k\)-paths, which cannot be extended to \((k\!+\!2)\)-path, so
\begin{align}
n_{k+2}\Big(\mathcal A_{k}\left(\mathbb{Z}^{d}\right)\Big)&=
\begin{array}{l}
\text{number of \((k\!+\!2)\)-paths to } \\
\mathbf v\in\mathcal A_{k}\left(\mathbb{Z}^{d}\right):\lVert\mathbf v\rVert=k\!+\!2
\end{array}
\le d^{k} \notag
\end{align}
A down-step, avoiding vertex repetition, extends \((k\!+\!2)\)-path to \((k\!+\!4)\)-path.
Two down-steps, avoiding vertex repetition, extend \(k\)-path to \((k\!+\!4)\)-path.
There are no more than \(2d^{k}\) \((k\!+\!4)\)-paths from \(\mathbf v_{0}\) to \(\mathcal A_{k}(\mathbb{Z}^{d})\), because other \((k\!+\!4)\)-paths would have to come from the extensions of non \((k\!+\!2)\)-paths and non \(k\)-paths, which is not possible.
Inductively, one down-step,.., and \(m\) down-steps, extend \((k\!+\!2m\!-\!2)\)-paths,.., and \(k\)-paths to \((k\!+\!2m)\)-paths.
There are no more than \(md^{k}\) \((k\!+\!2m)\)-paths from \(\mathbf v_{0}\) to \(\mathcal A_{k}(\mathbb{Z}^{d})\) or \(n_{k+2m}\left(\mathcal A_{k}\left(\mathbb{Z}^{d}\right)\right)\le md^{k}\), so
\begin{align}
\psi\left(\mathbb{Z}^{d},p\right)&=\lim_{k\to\infty}\displaystyle\sum_{i=0}n_{k+2i}\Big(\mathcal A_{k}\left(\mathbb{Z}^{d}\right)\Big)\cdot p^{k+2i} \notag \\
&\le\lim_{k\to\infty}(dp)^{k}\left(1+\displaystyle\sum_{i=1}i\cdot p^{2i}\right) \notag \\
\Rightarrow&\psi\left(\mathbb{Z}^{d},p<\displaystyle\frac{1}{d}\right)=0\Rightarrow p_{H}\left(\mathbb{Z}^{d}\right)\ge\displaystyle\frac{1}{d} \label{e2}
\end{align}
From inequalities (\ref{e1}) and (\ref{e2}),
\begin{align}
&\boxed{p_{H}\left(\mathbb{Z}^{d}\right)=\displaystyle\frac{1}{d}} \notag
\end{align}

\subsection{Modifications of \(\mathbb{Z}^{d}\)}
\noindent Graph \(\mathbf G\), as result of modification of \(\mathbb{Z}^{d}\), embedds \(\mathbb{Z}^{d}\) or modified \(\mathbb{Z}^{d}\).
Vertices in arc \(\mathcal A_{k}(\mathbf G)\) are \(k\) edges away from the origin \(\mathbf v_{0}\!\in\!\mathbf G\) and \(\mathcal A_{k}(\mathbf G)\) contains the vertices that belong to arcs \(\mathcal A_{k+i}(\mathbb{Z}^{d})\), which are visited in the shortest up-step traversal of \(\mathbf G\).
The shortest traversal to \(\mathcal A_{k}(\mathbf G)\) is traversal via neighborhoods \(\mathcal N_{u}(\mathbf v,\mathbf G)\), which builds \(k\)-paths ending in arcs \(\mathcal A_{k+i}(\mathbb{Z}^{d})\):
\begin{align}
\mathbf G&=\text{graph after modification of }\mathbb{Z}^{d} \notag \\
\mathcal A_{k}(\mathbf G)&=\!\!\!\bigcup_{\mathbf v\in\mathcal A_{k-1}(\mathbf G)}\!\!\!\mathcal N_{u}(\mathbf v,\mathbf G):\mathcal A_{1}(\mathbf G)=\mathcal B(\mathbf G) \notag \\
\mathcal N_{u}(\mathbf v,\mathbf G)&=\mathbf v\text{'s neighbors one edge closer to }\mathcal A_{k}(\mathbf G) \notag
\end{align}
Modification of \(\mathbb{Z}^{d}\) either removes or adds the edges from \(\mathcal B(\mathbb{Z}^{d})\) to some vertices in \(\mathbb{Z}^{d}\), so \(\mathcal B(\mathbf G)\) is composed of combinations of edges in \(\mathcal B(\mathbb{Z}^{d})\):
\begin{align}
\mathcal B(\mathbf G)&=\text{basis of }\mathbf G=\!\!\bigcup_{\forall \mathbf v\in\mathbf G}\!\!(\mathbf v-\mathbf v_{x}):\mathbf v_{x}\!\in\!\mathcal N(\mathbf v,\mathbf G) \notag
\end{align}
For all the rotations of \(\mathbf G\), there is the greatest number of \(k\)-paths to some \(\mathcal A_{k+i'}(\mathbb{Z}^{d})\), which minimizes percolating probability:
\begin{align}
n_{k}\Big(\mathcal A_{k+i}\left(\mathbb{Z}^{d}\right),\mathbf G\Big)&=
\begin{array}{l}
\text{number of \(k\)-paths ending in } \\
\mathcal A_{k+i}\left(\mathbb{Z}^{d}\right)\text{ in traversal of }\mathbf G
\end{array}
\notag \\
n_{k}\Big(\mathcal A_{k}(\mathbf G)\Big)&=
\begin{array}{l}
\text{number of \(k\)-paths to }\mathcal A_{k}(\mathbf G): \\
\displaystyle\sum_{i}n_{k}\Big(\mathcal A_{k+i}\left(\mathbb{Z}^{d}\right),\mathbf G\Big)
\end{array}
\notag
\end{align}
\(n_{k}\left(\mathcal A_{k+i}\left(\mathbb{Z}^{d}\right),\mathbf G\right)\) values are maximized by choosing \(\mathcal N_{u}(\mathbf v,\mathbf G)\), so that in the up-step traversal the greatest number of vertices get closer to \(\mathcal A_{k}(\mathbf G)\) in all rotations of \(\mathbb{Z}^{d}\) embeded in \(\mathbf G\).
Extensions of \(k\)-paths in \(\mathbf G\) are extensions of \(k\)-paths in \(\mathbb{Z}^{d}\). Thus, the extensions do not lower \(p'\): \(n_{k}\Big(\mathcal A_{k+i'}\left(\mathbb{Z}^{d}\right),\mathbf G\Big)\cdot p'\ge1\), so \(p'=p_{H}(\mathbf G)\).

\subsubsection*{\(1^{st}\) Example: Triangular Lattice}
\begin{figure}
\definecolor{gry}{gray}{.9}
\begin{pspicture}(4,4.2)(0,0)

\psline[linewidth=6pt,linestyle=solid,linecolor=gry]{-}(-0.2,4)(4.2,4)
\psline[linewidth=6pt,linestyle=solid,linecolor=gry]{-}(0.3,3.5)(3.7,3.5)
\psline[linewidth=6pt,linestyle=solid,linecolor=gry]{-}(0.8,3)(3.2,3)
\psline[linewidth=6pt,linestyle=solid,linecolor=gry]{-}(1.3,2.5)(2.7,2.5)

\psline[linewidth=0.5pt,linestyle=solid]{-}(0,3)(1,4)
\psline[linewidth=0.5pt,linestyle=solid]{-}(0,2)(2,4)
\psline[linewidth=0.5pt,linestyle=solid]{-}(0,1)(3,4)
\psline[linewidth=0.5pt,linestyle=solid]{-}(0,0)(4,4)
\psline[linewidth=0.5pt,linestyle=solid]{-}(1,0)(4,3)
\psline[linewidth=0.5pt,linestyle=solid]{-}(2,0)(4,2)
\psline[linewidth=0.5pt,linestyle=solid]{-}(3,0)(4,1)

\psline[linewidth=0.5pt,linestyle=solid]{-}(0,1)(1,0)
\psline[linewidth=0.5pt,linestyle=solid]{-}(0,2)(2,0)
\psline[linewidth=0.5pt,linestyle=solid]{-}(0,3)(3,0)
\psline[linewidth=0.5pt,linestyle=solid]{-}(0,4)(4,0)
\psline[linewidth=0.5pt,linestyle=solid]{-}(1,4)(4,1)
\psline[linewidth=0.5pt,linestyle=solid]{-}(2,4)(4,2)
\psline[linewidth=0.5pt,linestyle=solid]{-}(3,4)(4,3)

\multiput(0,0)(0.5,0){9}{\psline[linewidth=0.5pt,linestyle=solid]{-}(0,0)(0,4)}

{\color{white}
\multiput(0,0)(2,0){2}{\circle*{0.15}} \put(3,0){\circle*{0.15}}
\put(2.5,0.5){\circle*{0.15}}
\multiput(1,1)(2,0){2}{\circle*{0.15}}
\multiput(1.5,1.5)(2,0){2}{\circle*{0.15}}
\multiput(0,2)(1,0){2}{\circle*{0.15}}
\multiput(2.5,2.5)(1,0){2}{\circle*{0.15}}
\multiput(1,3)(3,0){2}{\circle*{0.15}}
\put(2.5,3.5){\circle*{0.15}}
}
\multiput(0,0)(2,0){2}{\circle{0.15}} \put(3,0){\circle{0.15}}
\put(2.5,0.5){\circle{0.15}}
\multiput(1,1)(2,0){2}{\circle{0.15}}
\multiput(1.5,1.5)(2,0){2}{\circle{0.15}}
\multiput(0,2)(1,0){2}{\circle{0.15}}
\multiput(2.5,2.5)(1,0){2}{\circle{0.15}}
\multiput(1,3)(3,0){2}{\circle{0.15}}
\put(2.5,3.5){\circle{0.15}}

\multiput(1,0)(3,0){2}{\circle*{0.15}}
\multiput(0.5,0.5)(1,0){2}{\circle*{0.15}} \put(3.5,0.5){\circle*{0.15}}
\multiput(0,1)(2,0){3}{\circle*{0.15}}
\multiput(0.5,1.5)(2,0){2}{\circle*{0.15}}
\multiput(3,2)(1,0){2}{\circle*{0.15}}
\multiput(0.5,2.5)(1,0){2}{\circle*{0.15}}
\multiput(0,3)(3,0){2}{\circle*{0.15}} \put(2,3){\circle*{0.15}}
\multiput(0.5,3.5)(1,0){2}{\circle*{0.15}} \put(3.5,3.5){\circle*{0.15}}
\multiput(0,4)(1,0){5}{\circle*{0.15}}

\put(2,2){\circle*{0.2}}

\psline[linewidth=0.5pt,linestyle=dotted,linearc=0.45]{-}(0.2,3.8)(3.8,3.8)(3.8,0.2)(3.8,0.2)(0.2,0.2)(0.2,3.8)

{\footnotesize
\put(1.7,1.9){0}
\put(2.6,2.44){\(\mathbf 1\!\!\uparrow_{1}\)}
\put(3.1,2.94){\(\mathbf 2\!\!\uparrow_{1}\)}
\put(3.6,3.44){\(\mathbf 3\!\!\uparrow_{1}\)}
\put(2.55,1.41){\(\mathbf 1\!\!\downarrow_{2}\)}
\put(3.07,0.92){\(\mathbf 2\!\!\downarrow_{2}\)}
\put(3.56,0.42){\(\mathbf 3\!\!\downarrow_{2}\)}
\put(1.57,2.41){\(\mathbf 1\!\!\uparrow_{2}\)}
\put(1.11,2.91){\(\mathbf 2\!\!\uparrow_{2}\)}
\put(0.59,3.41){\(\mathbf 3\!\!\uparrow_{2}\)}
\put(1.6,1.45){\(\mathbf 1\!\!\downarrow_{1}\)}
\put(1.1,0.9){\(\mathbf 2\!\!\downarrow_{1}\)}
\put(0.6,0.4){\(\mathbf 3\!\!\downarrow_{1}\)}

\put(1.45,4.12){\(\mathbf 2\!\uparrow_{1}\!\!+\mathbf 2\!\uparrow_{2}\)}
}

\end{pspicture}
\definecolor{gry}{gray}{.9}
\begin{pspicture}(4,4.2)(0,0)

\psline[linewidth=6pt,linestyle=solid,linecolor=gry]{-}(-0.2,4)(4.2,4)
\psline[linewidth=6pt,linestyle=solid,linecolor=gry]{-}(0.3,3.5)(3.7,3.5)
\psline[linewidth=6pt,linestyle=solid,linecolor=gry]{-}(0.8,3)(3.2,3)
\psline[linewidth=6pt,linestyle=solid,linecolor=gry]{-}(1.3,2.5)(2.7,2.5)

\psline[linewidth=0.5pt,linestyle=solid]{-}(0,3)(1,4)
\psline[linewidth=0.5pt,linestyle=solid]{-}(0,2)(2,4)
\psline[linewidth=0.5pt,linestyle=solid]{-}(0,1)(3,4)
\psline[linewidth=0.5pt,linestyle=solid]{-}(0,0)(4,4)
\psline[linewidth=0.5pt,linestyle=solid]{-}(1,0)(4,3)
\psline[linewidth=0.5pt,linestyle=solid]{-}(2,0)(4,2)
\psline[linewidth=0.5pt,linestyle=solid]{-}(3,0)(4,1)

\psline[linewidth=0.5pt,linestyle=solid]{-}(0,1)(1,0)
\psline[linewidth=0.5pt,linestyle=solid]{-}(0,2)(2,0)
\psline[linewidth=0.5pt,linestyle=solid]{-}(0,3)(3,0)
\psline[linewidth=0.5pt,linestyle=solid]{-}(0,4)(4,0)
\psline[linewidth=0.5pt,linestyle=solid]{-}(1,4)(4,1)
\psline[linewidth=0.5pt,linestyle=solid]{-}(2,4)(4,2)
\psline[linewidth=0.5pt,linestyle=solid]{-}(3,4)(4,3)

\multiput(0,0)(0,0.5){9}{\psline[linewidth=0.5pt,linestyle=solid]{-}(0,0)(4,0)}

{\color{white}
\multiput(0,0)(2,0){2}{\circle*{0.15}} \put(3,0){\circle*{0.15}}
\put(2.5,0.5){\circle*{0.15}}
\multiput(1,1)(2,0){2}{\circle*{0.15}}
\multiput(1.5,1.5)(2,0){2}{\circle*{0.15}}
\multiput(0,2)(1,0){2}{\circle*{0.15}}
\multiput(2.5,2.5)(1,0){2}{\circle*{0.15}}
\multiput(1,3)(3,0){2}{\circle*{0.15}}
\put(2.5,3.5){\circle*{0.15}}
\multiput(0,4)(2,0){3}{\circle*{0.15}}
}
\multiput(0,0)(2,0){2}{\circle{0.15}} \put(3,0){\circle{0.15}}
\put(2.5,0.5){\circle{0.15}}
\multiput(1,1)(2,0){2}{\circle{0.15}}
\multiput(1.5,1.5)(2,0){2}{\circle{0.15}}
\multiput(0,2)(1,0){2}{\circle{0.15}}
\multiput(2.5,2.5)(1,0){2}{\circle{0.15}}
\multiput(1,3)(3,0){2}{\circle{0.15}}
\put(2.5,3.5){\circle{0.15}}
\multiput(0,4)(2,0){3}{\circle{0.15}}

\multiput(1,0)(3,0){2}{\circle*{0.15}}
\multiput(0.5,0.5)(1,0){2}{\circle*{0.15}} \put(3.5,0.5){\circle*{0.15}}
\multiput(0,1)(2,0){3}{\circle*{0.15}}
\multiput(0.5,1.5)(2,0){2}{\circle*{0.15}}
\multiput(3,2)(1,0){2}{\circle*{0.15}}
\multiput(0.5,2.5)(1,0){2}{\circle*{0.15}}
\multiput(0,3)(3,0){2}{\circle*{0.15}} \put(2,3){\circle*{0.15}}
\multiput(0.5,3.5)(1,0){2}{\circle*{0.15}} \put(3.5,3.5){\circle*{0.15}}
\multiput(1,4)(2,0){2}{\circle*{0.15}}

\put(2,2){\circle*{0.2}}

\psline[linewidth=0.5pt,linestyle=dotted,linearc=0.45]{-}(0.2,3.8)(3.8,3.8)(3.8,0.2)(3.8,0.2)(0.2,0.2)(0.2,3.8)

{\footnotesize
\put(1.93,2.17){0}
\put(2.26,2.77){\(\mathbf 1\!\!\uparrow_{1}\)}
\put(2.76,3.27){\(\mathbf 2\!\!\uparrow_{1}\)}
\put(3.26,3.77){\(\mathbf 3\!\!\uparrow_{1}\)}
\put(2.26,1.77){\(\mathbf 1\!\!\downarrow_{2}\)}
\put(2.76,1.27){\(\mathbf 2\!\!\downarrow_{2}\)}
\put(3.26,0.77){\(\mathbf 3\!\!\downarrow_{2}\)}
\put(1.26,2.77){\(\mathbf 1\!\!\uparrow_{2}\)}
\put(0.76,3.27){\(\mathbf 2\!\!\uparrow_{2}\)}
\put(0.26,3.77){\(\mathbf 3\!\!\uparrow_{2}\)}
\put(1.26,1.77){\(\mathbf 1\!\!\downarrow_{1}\)}
\put(0.76,1.27){\(\mathbf 2\!\!\downarrow_{1}\)}
\put(0.26,0.77){\(\mathbf 3\!\!\downarrow_{1}\)}

\put(1.45,4.12){\(\mathbf 2\!\uparrow_{1}\!\!+\mathbf 2\!\uparrow_{2}\)}
}

\end{pspicture}
\caption{\label{latticeTexample}
Triangular lattice is built by adding \(\uparrow_{1}\!\!+\!\!\uparrow_{2}\) and \(\downarrow_{1}\!\!+\!\!\downarrow_{2}\)
(left) or by adding \(\uparrow_{1}\!\!+\!\!\downarrow_{2}\) and \(\downarrow_{1}\!\!+\!\!\uparrow_{2}\) (right) to each \(\mathbf v\in\mathbb{Z}^{2}\).
}
\end{figure}
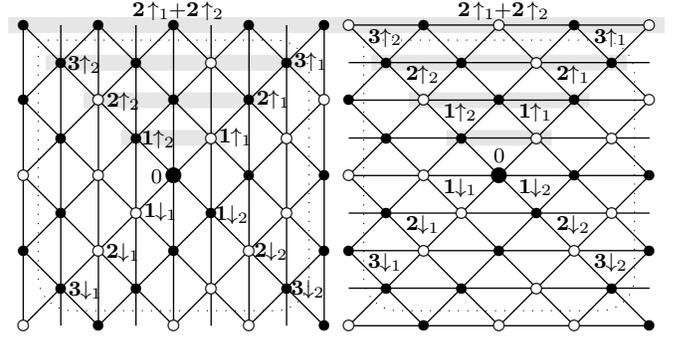

\noindent
Triangular lattice \(\mathbf T\) embeds \(\mathbb{Z}^{2}\), so it has two pairs of opposite sides, Fig. \ref{latticeTexample}.
\(n_{k}\left(\mathcal A_{k+i}\left(\mathbb{Z}^{d}\right),\mathbf T\right)\) values are maximized when each \(\mathbf v\!\in\!\mathbf T\) has neighbor \(\uparrow_{1}\!\!+\!\!\uparrow_{2}\).
\(\mathcal A_{k}(\mathbf T)\), \(k\) edges away from \(\mathbf v_{0}\!\in\!\mathbf T\), contains the vertices in arcs \(\mathcal A_{k+i}(\mathbb{Z}^{d})\) traversed via \(\mathcal N_{u}(\mathbf v,\mathbf T)\):
\begin{align}
\mathbf T&=\text{triangular lattice with embedded }\mathbb{Z}^{2} \notag \\
\mathcal B(\mathbf T)&=\{\uparrow_{1},\uparrow_{2},\uparrow_{1}\!\!+\!\!\uparrow_{2}\} \notag \\
\mathcal N_{u}(\mathbf v,\mathbf T)&=\mathbf v+\{\uparrow_{1},\uparrow_{2},\uparrow_{1}\!\!+\!\!\uparrow_{2}\} \notag \\
\mathcal A_{k}(\mathbf T)&\!=\!\!\!\!\!\!\!\!\displaystyle\bigcup_{\mathbf v\in\mathcal A_{k-1}(\mathbf T)}\!\!\!\!\!\!\!\mathcal N_{u}(\mathbf v,\mathbf T):\mathcal A_{1}(\mathbf T)\!=\!\{\uparrow_{1},\!\uparrow_{2},\!\uparrow_{1}\!\!+\!\!\uparrow_{2}\} \notag
\end{align}
\\
After \(1^{st}\) up-step in the traversal of \(\mathbf T\), there are 2 paths that end in \(\mathcal A_{1}(\mathbb{Z}^{2})\) and 1 path that ends in \(\mathcal A_{2}(\mathbb{Z}^{2})\):
\begin{align}
n_{1}\!\Big(\mathcal A_{1}\!\left(\mathbb{Z}^{2}\right)\!,\!\mathbf T\Big)&=n_{1}(1)=
\begin{array}{l}
\text{number of 1-paths to} \\
\mathcal A_{1}\!\left(\mathbb{Z}^{2}\right)\text{ traversing }\mathbf T
\end{array}=1\cdot2^{1}
\notag \\
n_{1}\!\Big(\mathcal A_{2}\!\left(\mathbb{Z}^{2}\right)\!,\!\mathbf T\Big)&=n_{1}(2)=
\begin{array}{l}
\text{number of 1-paths to} \\
\mathcal A_{2}\!\left(\mathbb{Z}^{2}\right)\text{ traversing }\mathbf T
\end{array}=1\cdot2^{0}
\notag \\
n_{1}\!\Big(\mathcal A_{1}\!\left(\mathbf T\right)\Big)&=n_{1}\Big(\mathcal A_{1}\left(\mathbb{Z}^{2}\right),\mathbf T\Big)+n_{1}\Big(\mathcal A_{2}\left(\mathbb{Z}^{2}\right),\mathbf T\Big) \notag
\end{align}
After \(2^{nd}\) up-step in the traversal of \(\mathbf T\), the number of paths doubles for the up-steps from \(\mathcal A_{1}(\mathbb{Z}^{2})\) to \(\mathcal A_{2}(\mathbb{Z}^{2})\) and from \(\mathcal A_{2}(\mathbb{Z}^{2})\) to \(\mathcal A_{3}(\mathbb{Z}^{2})\).
The number of paths does not change for the up-steps from \(\mathcal A_{1}(\mathbb{Z}^{2})\) to \(\mathcal A_{3}(\mathbb{Z}^{2})\) and from \(\mathcal A_{2}(\mathbb{Z}^{2})\) to \(\mathcal A_{4}(\mathbb{Z}^{2})\):
\begin{align}
n_{2}(2)&=2\cdot n_{1}(1)=1\cdot2^{2} \notag \\
n_{2}(3)&=n_{1}(1)+2\cdot n_{1}(2)=2\cdot2^{1} \notag \\
n_{2}(4)&=n_{1}(2)=1\cdot 2^{0} \notag
\end{align}
After \(k^{th}\) up-step in the traversal of \(\mathbf T\), the paths from \(\mathbf v_{0}\) end in \(\mathcal A_{k}(\mathbb{Z}^{2})\), \(\mathcal A_{k+1}(\mathbb{Z}^{2})\), .., \(\mathcal A_{2k}(\mathbb{Z}^{2})\):
\begin{align}
n_{k}(k)&=2\cdot n_{k-1}(k\!-\!1)=\displaystyle{k\choose0}\cdot2^{k} \notag \\
n_{k}(k\!+\!1)&=n_{k-1}(k\!-\!1)+2\cdot n_{k-1}(k)=\displaystyle{k\choose1}\cdot2^{k-1} \notag \\
n_{k}(k\!+\!2)&=n_{k-1}(k)+2\cdot n_{k-1}(k\!+\!1)=\displaystyle{k\choose2}\cdot2^{k-2} \notag \\
& .. \notag\\ 
n_{k}(2k\!-\!1)&=n_{k-1}(2k\!-\!3)\!+2\cdot n_{k-1}(2k\!-\!2)\!=\!\!\displaystyle{k\choose k-1}\cdot2^{1} \notag \\
n_{k}(2k)&=n_{k-1}(2k\!-\!2)=\displaystyle{k\choose k}\cdot2^{0} \notag
\end{align}
Inductively, after \((k\!+\!1)^{th}\) step
\begin{align}
n_{k+1}(k\!+\!1)&=2\cdot n_{k}(k) \notag \\
n_{k+1}(k\!+\!2)&=n_{k}(k)+2\cdot n_{k}(k\!+\!1) \notag \\
& .. \notag \\
n_{k+1}(2k\!+\!2)&=n_{k}(2k) \notag \\
&\Downarrow \notag \\
n_{k+1}(k\!+\!1)&=2\displaystyle{k\choose0}2^{k}=\displaystyle{k+1\choose0}\cdot2^{k+1} \notag \\
n_{k+1}(k\!+\!2)&=\displaystyle{k\choose0}2^{k}+2\displaystyle{k\choose1}2^{k-1}=\displaystyle{k+1\choose1}\cdot2^{k} \notag \\
& .. \notag \\
n_{k+1}(2k\!+\!2)&=\displaystyle{k\choose k}2^{0}=\displaystyle{k+1\choose k+1}\cdot2^{0} \notag
\end{align}
\begin{align}
n_{k}\Big(\mathcal A_{k}(\mathbf T)\Big)&=\displaystyle\sum_{i=0}^{i=k}{k\choose i}\cdot2^{k-i} \notag \\
n_{k}\Big(\mathcal A_{k+i}\left(\mathbb{Z}^{2}\right),\mathbf T\Big)&={k\choose i}\cdot2^{k-i} \notag
\end{align}
In the up-step traversal and for odd or even \(k\to\infty\),
the greatest number of paths end in \(\mathcal A_{k+\frac{k-1}{2}}(\mathbb{Z}^{2})\) or \(\mathcal A_{k+\frac{k}{2}-1}(\mathbb{Z}^{2})\), respectively, and
\(p_{H}(\mathbf T)\le\) minimum \(p\) for which \(\displaystyle{k\choose i}\cdot2^{k-i}\cdot p^{k}\ge1\):
\begin{align}
n_{k}\Big(\mathcal A_{k+\frac{k-1}{2}}\left(\mathbb{Z}^{2}\right)\Big)&=\displaystyle\binom{k}{\frac{k-1}{2}}\cdot2^{k-\left(\frac{k-1}{2}\right)} \notag \\
&\text{or} \notag \\
n_{k}\Big(\mathcal A_{k+\frac{k}{2}-1}\left(\mathbb{Z}^{2}\right)\Big)&=\displaystyle\binom{k}{\frac{k}{2}-1}\cdot2^{k-\left(\frac{k}{2}-1\right)} \notag
\end{align}
\begin{align}
\psi(\mathbf T,p)&=\lim_{k\to\infty}\left\{
\begin{array}{l}
2^{\frac{1}{2}}\left(\displaystyle\binom{k}{\frac{k-1}{2}}^{\frac{1}{k}} \cdot 2^{\frac{1}{2}} \cdot p\right)^{k} \\ \\
\ \ \ \text{or} \\ \\
2\left(\displaystyle\binom{k}{\frac{k}{2}-1}^{\frac{1}{k}} \cdot 2^{\frac{1}{2}} \cdot p\right)^{k}
\end{array}
\right.\notag \\
\Rightarrow p_{H}(\mathbf T)&=\frac{1}{2^{\frac{1}{2}}}\cdot\lim_{k\to\infty}\left\{
\begin{array}{l}
\displaystyle\frac{1}{\displaystyle\binom{k}{\frac{k-1}{2}}^{\frac{1}{k}}} \\ \\
\ \ \ \text{or} \\ \\
\displaystyle\frac{1}{\displaystyle\binom{k}{\frac{k}{2}\!-\!1}^{\frac{1}{k}}}
\end{array}
\right.
\approx\displaystyle\frac{1}{2^{\frac{3}{2}}}\approx0.3535 \notag
\end{align}
\(\lim_{k\to\infty}\displaystyle\binom{k}{\frac{k-1}{2}}^{\frac{1}{k}}\approx\lim_{k\to\infty}\displaystyle\binom{k}{\frac{k}{2}-1}^{\frac{1}{k}}\approx2\), \footnote{Limits at http://www.wolframalpha.com.}.

\subsubsection*{\(2^{nd}\) Example: Hexagonal Lattice}
\noindent Hexagonal lattice \(\mathbf H\) embeds modified \(\mathbb{Z}^{2}\), since it is obtained by removing from every other vertex \(\mathbf v\!\in\!\mathbb{Z}^{2}\) its neighbor \(\mathbf v+\!\!\uparrow_{1}\).
For any rotation of \(\mathbf H\), the neighborhood of \(\mathbf v\!\in\!\mathbf H\) is the same:
\begin{align}
\mathbf H&=\text{hexagonal lattice with modified }\mathbb{Z}^{2} \notag \\
\mathcal N_{u}(\mathbf v,\mathbf H)&=\mathbf v+\left\{
\begin{array}{ll}
\begin{array}{lr}
&\ \ \ \uparrow_{2}
\end{array}
&\text{ if }\mathbf v\!\in\!\mathcal A_{k}\left(\mathbb{Z}^{2}\right):k\text{ even}
\\ \\
\begin{array}{ll}
\uparrow_{1}& \\
&\uparrow_{2}
\end{array}
&\text{ if }\mathbf v\!\in\!\mathcal A_{k}\left(\mathbb{Z}^{2}\right):k\text{ odd}
\end{array}
\right\}
\notag
\end{align}
After \(k\) odd up-steps, \(2^{\frac{k+1}{2}}\) paths end in \(\mathcal A_{k}(\mathbb{Z}^{2})\).
\(\mathcal A_{1}(\mathbf H)=\{\uparrow_{1},\uparrow_{2}\}\) and
\begin{align}
n_{1}\Big(\mathcal A_{1}\left(\mathbb{Z}^{2}\right),\mathbf H\Big)&=2^{1} \notag \\
n_{2}\Big(\mathcal A_{2}\left(\mathbb{Z}^{2}\right),\mathbf H\Big)&=1\cdot n_{1}\Big(\mathcal A_{1}\left(\mathbb{Z}^{2}\right),\mathbf H\Big)=2^{1} \notag \\
n_{3}\Big(\mathcal A_{3}\left(\mathbb{Z}^{2}\right),\mathbf H\Big)&=2\cdot n_{2}\Big(\mathcal A_{2}\left(\mathbb{Z}^{2}\right),\mathbf H\Big)=2^{2} \notag \\
.. \notag \\
\Rightarrow
n_{k}\Big(\mathcal A_{k}\left(\mathbb{Z}^{2}\right),\mathbf H\Big)&=\left\{
\begin{array}{l}
2^{\frac{k+1}{2}}:k\text{ odd} \\
\ \ \ \text{or} \\
2^{\frac{k}{2}}:k\text{ even}
\end{array}
\right. \notag
\end{align}
When \(k\to\infty\),
\begin{align}
\psi(\mathbf H,p)=&\lim_{k\to\infty}\left\{
\begin{array}{l}
2^{\frac{k+1}{2}}\cdot p^{k}=2^{\frac{1}{2}}\left(2^{\frac{1}{2}}\cdot p\right)^{k} \\
\text{or} \\
2^{\frac{k}{2}}\cdot p^{k}=\left(2^{\frac{1}{2}}\cdot p\right)^{k}
\end{array}
\right.
\notag \\
\Rightarrow&p_{H}(\mathbf H)=\displaystyle\frac{1}{2^{\frac{1}{2}}}\approx0.7071 \notag
\end{align}

\section{Bond Percolation on \(\mathbb{Z}^{d}\)}
\begin{figure}
\definecolor{gry}{gray}{.95}
\definecolor{gry2}{gray}{.75}
\begin{pspicture}(8,8)(0,0)



\multiput(-0.1,1.45)(0,2.55){3}{\psline[linewidth=4pt,linestyle=solid,linecolor=gry]{-}(0,0)(8.1,0)}
\multiput(1.455,-0.1)(2.5,0){3}{\psline[linewidth=4pt,linestyle=solid,linecolor=gry]{-}(0,0)(0,8.1)}

\multiput(1.45,1.45)(2.5,0){3}{\pscircle[fillcolor=gry,fillstyle=solid,linestyle=none](0,0){0.2}}
\multiput(1.45,4)(2.5,0){3}{\pscircle[fillcolor=gry,fillstyle=solid,linestyle=none](0,0){0.2}}
\multiput(1.45,6.55)(2.5,0){3}{\pscircle[fillcolor=gry,fillstyle=solid,linestyle=none](0,0){0.2}}


\psline[linewidth=0.1pt,linestyle=solid]{-}(1.3,0.05)(7.9,6.7)
\psline[linewidth=0.1pt,linestyle=solid]{-}(3.85,0.05)(7.9,4.2)
\psline[linewidth=0.1pt,linestyle=solid]{-}(6.35,0.08)(7.9,1.6)
\psline[linewidth=0.1pt,linestyle=solid]{-}(1.7,8)(0,6.4)
\psline[linewidth=0.1pt,linestyle=solid]{-}(4.1,7.9)(0,3.8)
\psline[linewidth=0.1pt,linestyle=solid]{-}(6.7,8)(0,1.2)

\psline[linewidth=0.1pt,linestyle=solid]{-}(1.7,0)(0,1.7)
\psline[linewidth=0.1pt,linestyle=solid]{-}(4.2,0)(0,4.2)
\psline[linewidth=0.1pt,linestyle=solid]{-}(6.7,0)(0,6.7)
\psline[linewidth=0.1pt,linestyle=solid]{-}(1.1,8.1)(8.1,1.1)
\psline[linewidth=0.1pt,linestyle=solid]{-}(3.6,8.1)(8.1,3.6)
\psline[linewidth=0.1pt,linestyle=solid]{-}(6.1,8.1)(8.1,6.1)

\multiput(1.45,-0.03)(2.5,0){3}{\psline[linewidth=0.1pt,linestyle=solid]{-}(0,0)(0,8)}
\multiput(-0.02,1.45)(0,2.55){3}{\psline[linewidth=0.1pt,linestyle=solid]{-}(0,0)(8,0)}

\multiput(1.45,0.2)(2.52,0){3}{\circle*{0.17}}
\multiput(0.23,1.43)(2.5,0){4}{\circle*{0.17}}
\multiput(1.46,2.7)(2.5,0){3}{\circle*{0.17}}
\multiput(0.21,4)(2.5,0){4}{\circle*{0.17}}
\multiput(1.44,5.21)(2.52,0){3}{\circle*{0.17}}
\multiput(0.19,6.52)(2.5,0){4}{\circle*{0.17}}
\multiput(1.44,7.74)(2.5,0){3}{\circle*{0.17}}

\put(4.2,0.1){0}
\put(-0.05,4.3){\(\mathbf 3\!\!\uparrow_{2}\)}
\put(2.45,1.8){\(\mathbf 1\!\!\uparrow_{2}\)} \put(5,1.8){\(\mathbf 1\!\!\uparrow_{1}\)}

\put(4.1,4.2){\(\color{gry2}\mathbf v_{j}^{'}\)} \put(4.1,1.65){\color{gry2}\(\mathbf v_{i}^{'}\)}

{\footnotesize
\put(4.6,7.15){\(\mathbf 3\!\!\uparrow_{1}+\mathbf 2\!\!\uparrow_{2}\)}
\put(2.14,4.62){\(\mathbf 1\!\!\uparrow_{1}\!\!+\mathbf 2\!\!\uparrow_{2}\)}
\put(4.64,4.62){\(\mathbf 2\!\!\uparrow_{1}\!\!+\mathbf 1\!\!\uparrow_{2}\)}
}

\end{pspicture}
\caption{\label{siteToBondLattice}
Bond percolation on \(\mathbb{Z}^{2}\) (gray) is equivalent to site percolation on \(\mathbf B_{2}\) (black).
An edge \(\mathbf v_{i}^{'}\mathbf v_{j}^{'}=\uparrow_{2}^{'}\in\mathbb{Z}^{2}\) is a vertex \(\uparrow_{1}\!\!+\!\!\uparrow_{2}\in\mathbf B_{2}\).
\(\mp\!\uparrow_{2}^{'}\) connects \(\uparrow_{1}\!\!+\!\!\uparrow_{2}\) to its 2 neighbors: \(\uparrow_{1}\!\!+\!\!\uparrow_{2}\!\!+\!\!\uparrow_{2}^{'}=2\uparrow_{1}\!\!+2\uparrow_{2}\) and \(\uparrow_{1}\!\!+\!\!\uparrow_{2}\!\!-\!\!\uparrow_{2}^{'}=\mathbf v_{0}\).
Edges \(\mp\!\uparrow_{1}\) and \(\mp\!\uparrow_{2}\) connect \(\uparrow_{1}\!\!+\!\!\uparrow_{2}\) to its 4 other neighbors \(\uparrow_{1}\), \(\uparrow_{2}\), \(2\!\uparrow_{1}\!\!+\!\!\uparrow_{2}\), and \(\uparrow_{1}\!\!+2\!\uparrow_{2}\).
}
\end{figure}
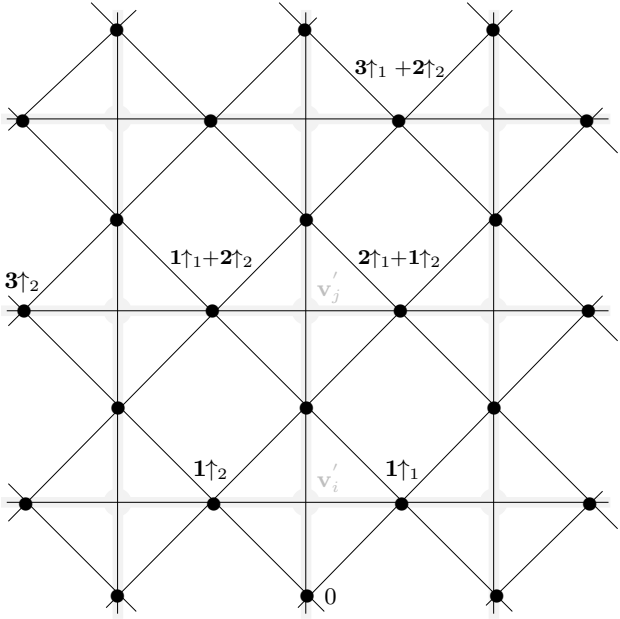

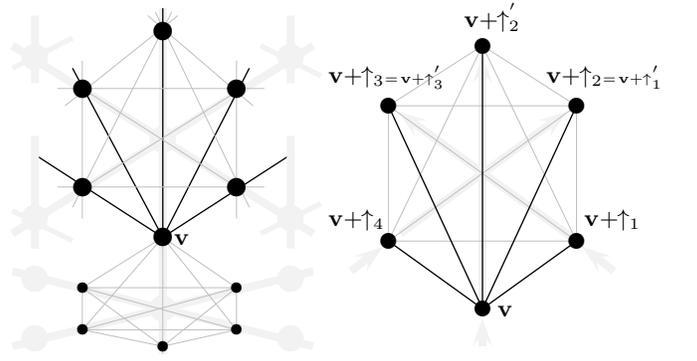
\begin{figure}
\definecolor{gry}{gray}{.95}
\definecolor{gry2}{gray}{.75}
\begin{pspicture}(8,4.5)(0,0)

{\color{gry}
\multiput(2,0.5)(0,2.25){2}{\circle*{0.35}}
\multiput(0.3,1.7)(0,2.15){2}{\circle*{0.35}} \multiput(3.7,1.7)(0,2.15){2}{\circle*{0.35}}
\multiput(0.3,0.1)(0,0.8){2}{\circle*{0.35}} \multiput(3.7,0.1)(0,0.8){2}{\circle*{0.35}}
}
\psline[linewidth=3pt,linecolor=gry]{-}(2,-0.1)(2,4.5)
\psline[linewidth=3pt,linecolor=gry]{-}(0,0)(4,1) \psline[linewidth=3pt,linecolor=gry]{-}(0,1)(4,0)
\psline[linewidth=3pt,linecolor=gry]{-}(0,1.5)(4,4) \psline[linewidth=3pt,linecolor=gry]{-}(0,4)(4,1.5)
\multiput(0,1.2)(3.3,2.2){2}{\psline[linewidth=2pt,linecolor=gry]{-}(0,0.6)(0.8,0.2)}
\multiput(0.3,3.5)(3.4,0){2}{\psline[linewidth=3pt,linecolor=gry]{-}(0,0)(0,0.9)} 
\multiput(0,4.3)(3.3,-2.2){2}{\psline[linewidth=2pt,linecolor=gry]{-}(0,-0.6)(0.8,-0.2)}

\psline[linewidth=0.5pt]{-}(2,1.45)(2,4.5) \psline[linewidth=0.5pt]{-}(0.8,3.7)(2,1.45) \psline[linewidth=0.5pt]{-}(0.3,2.55)(2,1.45) \psline[linewidth=0.5pt]{-}(3.15,3.7)(2,1.45) \psline[linewidth=0.5pt]{-}(3.7,2.55)(2,1.45)

\multiput(0.3,1.3)(3.4,0){2}{\psline[linewidth=3pt,linecolor=gry]{-}(0,0)(0,1.5)}

\psline[linewidth=0.1pt,linecolor=gry2]{-}(2,1.45)(2,-0.1)
\psline[linewidth=0.1pt,linecolor=gry2]{-}(0.93,1.7)(0.93,3.8) \psline[linewidth=0.1pt,linecolor=gry2]{-}(2.98,3.8)(2.98,1.7)
\psline[linewidth=0.1pt,linecolor=gry2]{-}(0.6,3.43)(3.35,3.43) \psline[linewidth=0.1pt,linecolor=gry2]{-}(0.6,2.12)(3.35,2.12)
\psline[linewidth=0.1pt,linecolor=gry2]{-}(0.93,3.43)(2.98,2.12) \psline[linewidth=0.1pt,linecolor=gry2]{-}(0.93,2.1)(3,3.4)
\psline[linewidth=0.1pt,linecolor=gry2]{-}(0.7,3.2)(2.2,4.4) \psline[linewidth=0.1pt,linecolor=gry2]{-}(0.8,1.85)(2.12,4.45)
\psline[linewidth=0.1pt,linecolor=gry2]{-}(3.2,3.2)(1.8,4.4) \psline[linewidth=0.1pt,linecolor=gry2]{-}(3.15,1.85)(1.9,4.45)

\psline[linewidth=0.1pt,linecolor=gry2]{-}(2.98,0.78)(2,1.45)
\psline[linewidth=0.1pt,linecolor=gry2]{-}(2.98,0.78)(0.93,0.78) \psline[linewidth=0.1pt,linecolor=gry2]{-}(2.98,0.78)(0.93,0.22) \psline[linewidth=0.1pt,linecolor=gry2]{-}(2.98,0.78)(2,0) \psline[linewidth=0.1pt,linecolor=gry2]{-}(2.98,0.78)(2.98,0.22)
\psline[linewidth=0.1pt,linecolor=gry2]{-}(0.93,0.78)(2,1.45) \psline[linewidth=0.1pt,linecolor=gry2]{-}(0.93,0.78)(0.93,0.22) \psline[linewidth=0.1pt,linecolor=gry2]{-}(0.93,0.78)(2.98,0.22) \psline[linewidth=0.1pt,linecolor=gry2]{-}(0.93,0.78)(2,0)
\psline[linewidth=0.1pt,linecolor=gry2]{-}(0.93,0.22)(2,1.45) \psline[linewidth=0.1pt,linecolor=gry2]{-}(0.93,0.22)(2,0) \psline[linewidth=0.1pt,linecolor=gry2]{-}(0.93,0.22)(2.98,0.22)
\psline[linewidth=0.1pt,linecolor=gry2]{-}(2,0)(2.98,0.22) \psline[linewidth=0.1pt,linecolor=gry2]{-}(2.98,0.22)(2,1.45)

\multiput(0.93,3.43)(2.05,0){2}{\circle*{0.25}} \multiput(0.93,2.12)(2.05,0){2}{\circle*{0.25}} \multiput(2,1.45)(0,2.75){2}{\circle*{0.25}}
\multiput(0.93,0.78)(2.05,0){2}{\circle*{0.15}}
\multiput(0.93,0.22)(2.05,0){2}{\circle*{0.15}}
\put(2,0){\circle*{0.15}}

\put(2.15,1.35){\(\mathbf v\)}

\psline[linewidth=3pt,linecolor=gry]{->}(6.25,0)(6.25,0.37) \psline[linewidth=3pt,linecolor=gry]{->}(6.25,0.6)(6.25,3.87)
\psline[linewidth=3pt,linecolor=gry]{->}(8,1)(7.58,1.32) \psline[linewidth=3pt,linecolor=gry]{->}(7.44,1.45)(5.13,3.12)
\psline[linewidth=3pt,linecolor=gry]{->}(4.5,1)(4.92,1.32) \psline[linewidth=3pt,linecolor=gry]{->}(5.16,1.5)(7.37,3.12)

\psline[linewidth=0.1pt,linecolor=gry2]{-}(5,1.4)(5,3.2) \psline[linewidth=0.1pt,linecolor=gry2]{-}(5,1.4)(6.25,4) \psline[linewidth=0.1pt,linecolor=gry2]{-}(5,1.4)(7.5,3.2) \psline[linewidth=0.1pt,linecolor=gry2]{-}(5,1.4)(7.5,1.4)
\psline[linewidth=0.1pt,linecolor=gry2]{-}(5,3.2)(6.25,4) \psline[linewidth=0.1pt,linecolor=gry2]{-}(5,3.2)(7.5,3.2) \psline[linewidth=0.1pt,linecolor=gry2]{-}(5,3.2)(7.5,1.4)
\psline[linewidth=0.1pt,linecolor=gry2]{-}(6.25,4)(7.5,3.2) \psline[linewidth=0.1pt,linecolor=gry2]{-}(6.25,4)(7.5,1.4)
\psline[linewidth=0.1pt,linecolor=gry2]{-}(7.5,3.2)(7.5,1.4)

\multiput(5,3.2)(2.5,0){2}{\circle*{0.2}} \multiput(5,1.4)(2.5,0){2}{\circle*{0.2}} \multiput(6.25,0.5)(0,3.5){2}{\circle*{0.2}}
\psline[linewidth=0.5pt]{-}(6.25,0.5)(6.25,4) \psline[linewidth=0.5pt]{-}(6.25,0.5)(5,3.2) \psline[linewidth=0.5pt]{-}(6.25,0.5)(5,1.4) \psline[linewidth=0.5pt]{-}(6.25,0.5)(7.5,3.2) \psline[linewidth=0.5pt]{-}(6.25,0.5)(7.5,1.4)

\put(6.45,0.4){\(\mathbf v\)} \put(4.2,1.6){\(\mathbf v+\!\!\uparrow_{4}\)} \put(4.2,3.5){\(\mathbf v+\!\!\uparrow_{3}\)\tiny\(=\!\mathbf v+\!\!\uparrow_{3}^{'}\)} \put(6,4.25){\(\mathbf v+\!\!\uparrow_{2}^{'}\)} \put(7.1,3.5){\(\mathbf v+\!\!\uparrow_{2}\)\tiny\(=\!\mathbf v+\!\!\uparrow_{1}^{'}\)} \put(7.6,1.6){\(\mathbf v+\!\!\uparrow_{1}\)}

\end{pspicture}
\caption{\label{latticesEG6}
2-dimensional projection of neighborhoods \(\mathcal N(\mathbf v,\mathbf B_{3})\) (left) and \(\mathcal N_{u}(\mathbf v,\mathbf B_{3})\) (right).
\(\mathbf v\in\mathcal A_{k}\left(\mathbb{Z}^{2\cdot3-2})=\mathbb{Z}^{4}\right)\) has 1 up-step neighbor \(\mathbf v+\!\!\uparrow_{2}^{'}\) in arc \(\mathcal A_{k+2}(\mathbb{Z}^{4})\) and 4 up-step neighbors \(\mathbf v+\!\!\uparrow_{i}\) in arc \(\mathcal A_{k+1}(\mathbb{Z}^{4})\).
Up-step neighbors of \(\mathbf v+\!\!\uparrow_{i}\) are in \(\mathcal A_{k+2}(\mathbb{Z}^{4})\) and traversing away from the origin.
Each \(\mathbf v+\!\!\uparrow_{i}\) has 2 up-step neighbors toward edges \(\uparrow_{2}^{'}\).
}
\end{figure}

\noindent
Bond percolation on \(\mathbb{Z}^{d}\) is equivalent to site percolation if the edges of \(\mathbb{Z}^{d}\) are translated into vertices, which are connected if the bonds in \(\mathbb{Z}^{d}\) shared a vertex, \citep{bollobas2006}:
\begin{align}
\mathbf B_{d}&=\text{translation of edges of }\mathbb{Z}^{d}\text{ to new vertices} \notag
\end{align}
A vertex in \(\mathbf B_{d}\) is connected to its neighbors via \(2\cdot(2d-1)\) edges: each end of edge in \(\mathbb{Z}^{d}\) meets with \((2d-1)\) edges, Fig. \ref{siteToBondLattice}.
An edge \(\mathbf v_{i}^{'}\mathbf v_{j}^{'}\) and its adjecent edge along the same dimension in \(\mathbb{Z}^{d}\) are translated into vertices \(\mathbf v_{i}\) and \(\mathbf v_{j}\) in \(\mathbf B_{d}\), so the edges \(\mathbf v_{i}^{'}\mathbf v_{j}^{'}\in\mathbb{Z}^{d}\) and \(\mathbf v_{i}\mathbf v_{j}\in\mathbf B_{d}\) are along the same dimension.
\(\mathbf v_{i}^{'}\mathbf v_{j}^{'}=\uparrow_{i}^{'}\) is in basis of \(\mathbb{Z}^{d}\) and \(\mp\!\uparrow_{i}^{'}\) connects \(\mathbf v_{i}\) to its up-step and down-step neighbors in \(\mathbf B_{d}\).
Other \(2\cdot(2d-2)\) edges connect \(\mathbf v_{i}\) to its \((2d-2)\) up-step and \((2d-2)\) down-step neighbors.
\\ \\
After the translation of \(\mathbb{Z}^{d}\), \(\mathbb{Z}^{d}\) and \(\mathbb{Z}^{2d-2}\) are embeded in \(\mathbf B_{d}\).
Each \(\mathbf v\in\mathbf B_{d}\) has neighbors \(\mathbf v+\mathbf u\): \(\mathbf u\in\mathcal B\left(\mathbb{Z}^{2d-d}\right)\) and every other \(\mathbf v\in\mathbf B_{d}\) has neighbors \(\mathbf v+\mathbf t\): \(\mathbf t\in\mathcal B\left(\mathbb{Z}^{d}\right)\):
\begin{align}
\mathbb{Z}^{2d-2}\!=&\text{lattice embeded in }\mathbf B_{d}\text{ after translation of }\mathbb{Z}^{d} \notag \\
\mathcal B(\mathbb{Z}^{d})\!=&\text{basis of }\mathbb{Z}^{d}=\{\uparrow_{1}^{'},..,\uparrow_{d}^{'}\} \notag \\
\mathcal B(\mathbb{Z}^{2d-2})\!=&\text{basis of }\mathbb{Z}^{2d-2}=\{\uparrow_{1},..,\uparrow_{2d-2}\} \notag \\
\mathcal B(\mathbf B_{d})\!=&\text{basis of }\mathbf B_{d}=\mathcal B(\mathbb{Z}^{d})\bigcup\mathcal B(\mathbb{Z}^{2d-2}) \notag
\end{align}
\(\mathbf B_{d}\) embeds \(\mathbb{Z}^{2d-2}\), so it has \((2d-2)\) pairs of opposite arcs, which in any rotation of \(\mathbf B_{d}\) look the same, Fig. \ref{latticesEG6}.

\subsection{Up-step Traversal of \(\mathbf B_{d}\)}
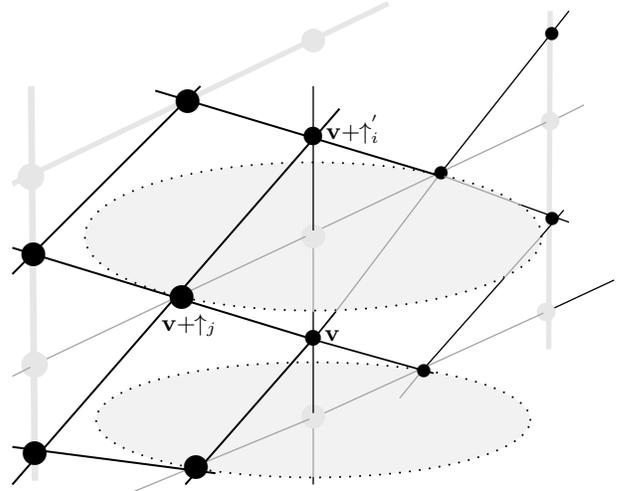
\begin{figure}
\definecolor{gry}{gray}{.95} \definecolor{gry2}{gray}{.65} \definecolor{gry3}{gray}{.90}
\begin{pspicture}(8,6)(0,0)

\psellipse[fillcolor=gry,fillstyle=solid,linestyle=dotted](4,3)(3.05,1)
\psellipse[fillcolor=gry,fillstyle=solid,linestyle=dotted](4,0.56)(2.9,0.78)

\psline[linewidth=0.5pt,linecolor=gry2]{-}(4,2)(4,3)
\psline[linewidth=0.5pt,linecolor=gry2]{-}(7.2,4.55)(7.6,4.75)
\put(7.15,4.54){\color{gry3}\circle*{0.26}}
\psline[linewidth=0.5pt,linecolor=gry2]{-}(4.1,3.1)(7.1,4.52)
\put(4,3){\color{gry3}\circle*{0.3}}

\psline[linewidth=0.5pt,linecolor=gry2]{-}(2.2,2.2)(3.9,3)
\psline[linewidth=0.5pt,linecolor=gry2]{-}(0.2,1.25)(2.2,2.2)
\psline[linewidth=0.5pt,linecolor=gry2]{-}(4.3,2)(5.7,3.85) 
\psline[linewidth=0.5pt,linecolor=gry2]{-}(4,0.6)(4,-0.2)

\psline[linewidth=2pt,linecolor=gry3]{-}(7.13,1.5)(7.15,6)
\psline[linewidth=0.5pt]{-}(7.1,2)(8,2.42)
\put(7.1,2){\color{gry3}\circle*{0.26}} 
\psline[linewidth=0.5pt,linecolor=gry2]{-}(4.08,0.62)(7.06,1.98)
\put(4,0.6){\color{gry3}\circle*{0.3}}
\psline[linewidth=0.5pt,linecolor=gry2]{-}(3.95,0.57)(1.6,-0.4)
\psline[linewidth=0.5pt]{-}(4,1.6)(4,0.65) 
\psline[linewidth=0.5pt]{-}(4,-0.2)(4,-0.3)
\psline[linewidth=2.5pt,linecolor=gry3]{-}(0.3,-0.25)(0.25,5)
\put(0.3,1.3){\color{gry3}\circle*{0.34}} 
\psline[linewidth=0.5pt,linecolor=gry2]{-}(0.24,1.26)(0,1.14)
\put(0.25,3.8){\color{gry3}\circle*{0.34}}
\put(0.28,2.76){\circle*{0.3}}

\psline[linewidth=2pt,linecolor=gry3]{-}(0,3.7)(5,6.1)
\put(4,5.6){\color{gry3}\circle*{0.3}}

\put(7.17,5.7){\circle*{0.16}}
\put(4,1.65){\circle*{0.22}} \put(4.15,1.62){\(\mathbf v\)}
\put(4,4.34){\circle*{0.25}} \put(4.17,4.3){\(\mathbf v+\!\!\uparrow_{i}^{'}\!\!\)}

\put(2.25,2.2){\circle*{0.3}} \put(1.98,1.75){\(\mathbf v+\!\!\uparrow_{j}\)}
\psline[linewidth=0.65pt]{-}(4,1.63)(4.3,2)
\put(5.7,3.85){\circle*{0.18}} 
\psline[linewidth=0.5pt]{-}(5.7,3.85)(7.4,6)
\psline[linewidth=0.5pt,linecolor=gry2]{-}(5.74,3.8)(6.76,3.42)
\psline[linewidth=0.5pt]{-}(6.76,3.42)(7.4,3.19)
\put(7.18,3.24){\circle*{0.16}}

\psline[linewidth=0.5pt]{-}(4,1.6)(4,2)
\psline[linewidth=0.5pt]{-}(4,3.08)(4,5)
\put(0.3,0.12){\circle*{0.32}}
\put(2.44,-0.07){\circle*{0.3}}
\psline[linewidth=0.5pt,linecolor=gry2]{-}(5.47,1.24)(5.15,0.85) \put(5.47,1.22){\circle*{0.18}} \psline[linewidth=0.5pt]{-}(5.47,1.24)(6.54,2.45) \psline[linewidth=0.5pt,linecolor=gry2]{-}(6.54,2.45)(7.02,2.99) \psline[linewidth=0.5pt]{-}(7.02,2.99)(7.33,3.35)
\put(2.33,4.8){\circle*{0.3}}

\psline[linewidth=0.75pt]{-}(0,-0.3)(4.35,4.7)
\psline[linewidth=0.75pt]{-}(2.5,5)(0,2.5)
\psline[linewidth=0.75pt]{-}(2.7,-0.15)(0,0.2)
\psline[linewidth=0.75pt]{-}(4.1,1.6)(0,2.85) \psline[linewidth=0.65pt]{-}(4.1,1.6)(5.6,1.18)
\psline[linewidth=0.75pt]{-}(5.7,3.8)(1.9,4.94)
\psline[linewidth=0.75pt]{-}(2.3,-0.3)(4,1.64)

\end{pspicture}
\caption{\label{lattice7}
Plane parallel to edge \(\uparrow_{i}^{'}\in\mathcal B(\mathbb{Z}^{d})\) intersects all the other edges in \(\mathcal B(\mathbb{Z}^{d})\), which are squeezed to 2-dimensional plane (gray ellipse).
Each \(\mathbf v+\!\!\uparrow_{j}\) has only 2 up-step neighbors.
}
\end{figure}

\noindent The shortest up-step traversal to arc \(\mathcal A_{k}(\mathbf B_{d})\) is a traversal along edges \(\uparrow_{i}^{'}\) and arcs of \(\mathbb{Z}^{2d-2}\) embeded in \(\mathbf B_{d}\):
\begin{align}
\mathcal A_{k}(\mathbf B_{d})&=\text{vertices \(k\) up-steps away from }\mathbf v_{0}\in\mathbf B_{d} \notag \\
\mathcal A_{k}(\mathbb{Z}^{2d-2})&=\text{vertices \(k\) up-steps away from }\mathbf v_{0}\in\mathbb{Z}^{2d-2} \notag \\
\mathbf v_{0}\in\mathbf B_{d}&=\mathbf v_{0}\in\mathbb{Z}^{2d-2} \notag \\
\uparrow_{i}^{'}&=\text{edge toward }\mathcal A_{k}(\mathbf B_{d}):k\!\uparrow_{i}^{'}\in\mathcal A_{k}(\mathbf B_{d}) \notag
\end{align}
A vertex \(\mathbf v\) at the end of edge \(\uparrow_{i}^{'}\) has 1 up-step neighbor \(\mathbf v+\!\!\uparrow_{i}^{'}\) and \((2d\!-\!2)\) up-step neighbors \(\mathbf v+\!\!\uparrow_{j}\).
For \(\mathbf v\in\mathcal A_{k}(\mathbf B_{d})\) and \(\mathbf u\in\mathcal A_{k}(\mathbb{Z}^{2d-2})\)
\begin{align}
\left(\mathbf v+\!\!\uparrow_{i}^{'}\right)\in&\ \mathcal A_{k+1}(\mathbf B_{d})\ \&\ \left(\mathbf u+\!\!\uparrow_{i}^{'}\right)\in\mathcal A_{k+2}\left(\mathbb{Z}^{2d-2}\right) \notag \\
(\mathbf v+\!\!\uparrow_{j})\in&\ \mathcal A_{k+1}(\mathbf B_{d})\ \&\ (\mathbf u+\!\!\uparrow_{j})\in\mathcal A_{k+1}\left(\mathbb{Z}^{2d-2}\right) \notag
\end{align}
Vertex \(\mathbf v=\mathbf v_{x}+\!\!\uparrow_{i}^{'}\) is at \(\uparrow_{i}^{'}\) edge and it has 1 up-step neighbor \(\mathbf v+\!\!\uparrow_{i}^{'}\) at edge \(\uparrow_{i}^{'}\) and it has \((2d\!-\!2)\) up-step neighbors \(\mathbf v+\!\!\uparrow_{j}\), which are not at edge \(\uparrow_{i}^{'}\).
Since \(\mathbf v+\!\!\uparrow_{j}\) is not at edge \(\uparrow_{i}^{'}\), it connects to two \(\uparrow_{i}^{'}\) edges in the up-step traversal (and two \(\uparrow_{i}^{'}\) edges in the down-step traversal), Fig. \ref{lattice7}.
All the other connections from \(\mathbf v+\!\!\uparrow_{j}\) are to the non-\(\uparrow_{j}^{'}\) edges, which are either not in up-step traversal or already traversed via \(\mathbf v\).

\subsection{Critical Probability for Bond Percolation}
\noindent For any rotation of \(\mathbf B_{d}\), the definition of up-step neighborhoods is the same:
\begin{align}
\mathcal N_{u}(\mathbf v,\mathbf B_{d})&=\mathbf v+\left\{
\begin{array}{ll}
\uparrow_{j},\uparrow_{i}^{'}:\forall\uparrow_{j}&\text{ if }\mathbf v\text{ is on}\uparrow_{i}^{'} \\
\uparrow_{i_1},\uparrow_{i_2}&\text{ if }\mathbf v\text{ is not on}\uparrow_{i}^{'}
\end{array}
\right\}
\notag \\
&\uparrow_{j_1},\uparrow_{j_2},\uparrow_{j}\in\mathcal B(\mathbb{Z}^{2d-2}) \notag \\
&\uparrow_{i}^{'}\in\mathcal B(\mathbb{Z}^{d}) \notag \\
&\mathbf v+\!\!\uparrow_{j_1}\&\ \mathbf v+\!\!\uparrow_{j_2}\text{are on}\uparrow_{i}^{'}\text{edges} \notag \\
&\mathbf v+\!\!\uparrow_{j}\text{are not on}\uparrow_{i}^{'}\text{edges} \notag
\end{align}
Vertices of \(\mathbf B_{d}\) traversed at \(k^{th}\) up-step are in arc \(\mathcal A_{k}(\mathbf B_{d})\).
For \(\mathbf v_{0}\) at \(\uparrow_{i}^{'}\),
\begin{align}
\mathcal A_{1}(\mathbf B_{d})&=\left\{\uparrow_{i}^{'}\right\}\bigcup\left\{\bigcup_{j}\uparrow_{j}\right\}:\ \uparrow_{i}^{'}\in\!\mathcal B(\mathbb{Z}^{d}),\uparrow_{j}\in\!\mathcal B(\mathbb{Z}^{2d-2}) \notag \\
\mathcal A_{k}(\mathbf B_{d})&=\!\!\!\!\bigcup_{\mathbf v\in\mathcal A_{k-1}(\mathbf B_{d})}\!\!\!\!\mathcal N_{u}(\mathbf v,\mathbf B_{d}) \notag
\end{align}
From \(\mathbf v_{0}\), there is one up-step via edges in \(\mathcal B\left(\mathbb{Z}^{2d-2}\right)\) toward \(\mathcal A_{1}\left(\mathbb{Z}^{2d-2}\right)\subset\mathcal A_{1}\left(\mathbf B_{d}\right)\) and there is one up-step via edges in \(\mathcal B\left(\mathbb{Z}^{d}\right)\) toward \(\mathcal A_{2}\left(\mathbb{Z}^{2d-2}\right)\subset\mathcal A_{1}\left(\mathbf B_{d}\right)\):
\begin{align}
&\begin{array}{lll}
n_{1}\!\Big(\mathcal A_{1}\!\left(\mathbb{Z}^{2d-2}\right)\!,\mathbf B_{d}\Big)&\!\!\!=n_{1}(1)=&\!\!\!\!
\begin{array}{l}
\text{number of 1-paths to} \\
A_{1}\!\left(\mathbb{Z}^{2d-2}\right)\!\text{ traversing }\mathbf B_{d}
\end{array}
\\
n_{1}\!\Big(\mathcal A_{2}\!\left(\mathbb{Z}^{2d-2}\right)\!,\mathbf B_{d}\Big)&\!\!\!=n_{1}(2)=&\!\!\!\!
\begin{array}{l}
\text{number of 1-paths to} \\
A_{2}\!\left(\mathbb{Z}^{2d-2}\right)\!\text{ traversing }\mathbf B_{d}
\end{array}
\end{array}
\notag \\
&n_{1}\Big(\mathcal A_{1}\left(\mathbf B_{d}\right)\Big)=n_{1}(1)+n_{1}(2)=\displaystyle\sum_{i=1}^{2}n_{1}(i)=(2d-2)+1 \notag
\end{align}
For each vertex in \(\mathcal A_{1}(\mathbb{Z}^{2d-2})\), there are 2 up-step neighbors in \(\mathcal A_{2}(\mathbb{Z}^{2d-2})\).
For each vertex in \(\mathcal A_{2}(\mathbb{Z}^{2d-2})\), there are \((2d-2)\) up-step neighbors in \(\mathcal A_{3}(\mathbb{Z}^{2d-2})\) and 1 up-step neighbor in \(\mathcal A_{4}(\mathbb{Z}^{2d-2})\):
\begin{align}
n_{2}(2)&=2\cdot n_{1}(1)=2\cdot(2d-2) \notag \\
n_{2}(3)&=(2d-2)\cdot n_{1}(2)=(2d-2) \notag \\
n_{2}(4)&=1\cdot n_{1}(2)=1 \notag \\
n_{2}\Big(\mathcal A_{2}\left(\mathbf B_{d}\right)\Big)&=\displaystyle\sum_{i=2}^{4}n_{2}(i) \notag
\end{align}
Vertices in \(\mathcal A_{3}(\mathbf B_{d})\) are vertices of paths ending in arcs \(\mathcal A_{3}(\mathbb{Z}^{2d-2})\), \(\mathcal A_{4}(\mathbb{Z}^{2d-2})\), \(\mathcal A_{5}(\mathbb{Z}^{2d-2})\), and \(\mathcal A_{6}(\mathbb{Z}^{2d-2})\):
\begin{align}
n_{3}(3)&=(2d-2)\cdot n_{2}(2)=2\cdot (2d-2)^{2} \notag \\
n_{3}(4)&=n_{2}(2)+2\cdot n_{2}(3)=4\cdot(2d-2) \notag \\
n_{3}(5)&=(2d-2)\cdot n_{2}(4)=(2d-2) \notag \\
n_{3}(6)&=n_{2}(4)=1 \notag \\
n_{3}\Big(\mathcal A_{3}\left(\mathbf B_{d}\right)\Big)&=\displaystyle\sum_{i=3}^{6}n_{3}(i) \notag
\end{align}
In the traversal from a vertex in \(\mathcal A_{k}(\mathbb{Z}^{2d-d})\) to \(\mathcal A_{k+1}(\mathbb{Z}^{2d-d})\) the number of paths doubles, if \(k\) is odd.
If \(k\) is even, an up-step from a vertex in \(\mathcal A_{k}(\mathbb{Z}^{2d-d})\) to \(\mathcal A_{k+1}(\mathbb{Z}^{2d-d})\) \((2d\!-\!2)\)-tuples the number of paths, and an up-step from a vertex in \(\mathcal A_{k}(\mathbb{Z}^{2d-d})\) to \(\mathcal A_{k+2}(\mathbb{Z}^{2d-d})\) does not change the number of paths.
After \(k^{th}\) up-step, the paths end in \(\mathcal A_{k}(\mathbb{Z}^{2d-2})\), \(\mathcal A_{k+1}(\mathbb{Z}^{2d-2})\),.., \(\mathcal A_{2k}(\mathbb{Z}^{2d-2})\).
\\ \\
For \(D=(2d-2)\),
\begin{align}
\displaystyle\sum_{i=4}^{8}n_{4}(i)&=4D^{2}+4D^{2}+6D^{1}+D^{1}+1 \notag \\
\displaystyle\sum_{i=5}^{10}n_{5}(i)&=4D^{3}+12D^{2}+6D^{2}+8D^{1}+D^{1}+1 \notag \\
\displaystyle\sum_{i=6}^{12}n_{6}(i)&=8D^{3}+12D^{3}+24D^{2}+8D^{2}+10D^{1}+D^{1}+1 \notag \\
\displaystyle\sum_{i=7}^{14}n_{7}(i)&=8D^{4}+32D^{3}+24D^{3}+40D^{2}+10D^{2}+.. \notag \\
\displaystyle\sum_{i=8}^{16}n_{8}(i)&=16D^{4}+32D^{4}+80D^{3}+40D^{3}+60D^{2}+.. \notag \\
\displaystyle\sum_{i=9}^{18}n_{9}(i)&=16D^{5}+80D^{4}+80D^{4}+160D^{3}+60D^{3}+.. \notag \\
\displaystyle\sum_{i=10}^{20}n_{10}(i)&=32D^{5}+80D^{5}+240D^{4}+160D^{4}+280D^{3}+.. \notag \\
&.. \notag
\end{align}
For even \(k\),
\begin{align}
n_{k-1}\Big(\mathcal A_{k-1}\left(\mathbf B_{d}\right)\Big)&=\displaystyle\sum_{i=k-1}^{2k-2}n_{k-1}(i)=
\begin{array}{c}
2^{\frac{k-2}{2}}D^{\frac{k}{2}}\ + \\
\displaystyle{\frac{k}{2}\choose1}2^{\frac{k-2}{2}}D^{\frac{k-2}{2}}\ + \\
\displaystyle{\frac{k}{2}\choose2}2^{\frac{k-4}{2}}D^{\frac{k-2}{2}}\ + \\
\displaystyle{\frac{k+2}{2}\choose3}2^{\frac{k-4}{2}}D^{\frac{k-4}{2}}\ + \\
.. 
\end{array} \notag \\ \notag \\
n_{k}\Big(\mathcal A_{k}\left(\mathbf B_{d}\right)\Big)&=\displaystyle\sum_{i=k}^{2k}n_{k}(i)=
\begin{array}{c}
2^{\frac{k}{2}}D^{\frac{k}{2}}\ + \\
\displaystyle{\frac{k}{2}\choose1}2^{\frac{k-2}{2}}D^{\frac{k}{2}}\ + \\
\displaystyle{\frac{k+2}{2}\choose2}2^{\frac{k-2}{2}}D^{\frac{k-2}{2}}\ + \\
\displaystyle{\frac{k+2}{2}\choose3}2^{\frac{k-4}{2}}D^{\frac{k-2}{2}}\ + \\
..
\end{array} \notag
\end{align}
Inductively, for \((k+1)\) and \((k+2)\),
\begin{align}
\displaystyle\sum_{i=k+1}^{2k+2}n_{k+1}(i)&=
\begin{array}{c}
D\cdot n_{k}(k)\ + \\
1\cdot n_{k}(k)+2\cdot n_{k}(k+1)\ + \\
D\cdot n_{k}(k+2)\ +\\
1\cdot n_{k}(k+2)+2\cdot n_{k}(k+3)\ + \\
..
\end{array}
\notag \\ \notag \\
n_{k+1}\Big(\mathcal A_{k+1}\left(\mathbf B_{d}\right)\Big)&=\begin{array}{c}
2^{\frac{k}{2}}D^{\frac{k+2}{2}}\ + \\
\displaystyle{\frac{k+2}{2}\choose1}2^{\frac{k}{2}}D^{\frac{k}{2}} + \\
\displaystyle{\frac{k+2}{2}\choose2}2^{\frac{k-2}{2}}D^{\frac{k}{2}}\ + \\
\displaystyle{\frac{k+4}{2}\choose3}2^{\frac{k-2}{2}}D^{\frac{k-2}{2}}\ + \\
..
\end{array} \notag
\end{align}
\begin{align}
\displaystyle\sum_{i=k+2}^{2k+4}n_{k+2}(i)&=
\begin{array}{c}
2\cdot n_{k+1}(k+1)\ + \\
D\cdot n_{k+1}(k+2)\ +\\
1\cdot n_{k+1}(k+2)+2\cdot n_{k+1}(k+3)\ + \\
D\cdot n_{k+1}(k+4)\ +\\
..
\end{array}
\notag \\ \notag \\
n_{k+2}\Big(\mathcal A_{k+2}\left(\mathbf B_{d}\right)\Big)&=\begin{array}{c}
2^{\frac{k+2}{2}}D^{\frac{k+2}{2}}\ + \\
\displaystyle{\frac{k+2}{2}\choose1}2^{\frac{k}{2}}D^{\frac{k+2}{2}}\ + \\
\displaystyle{\frac{k+4}{2}\choose2}2^{\frac{k}{2}}D^{\frac{k}{2}}\ + \\
\displaystyle{\frac{k+4}{2}\choose3}2^{\frac{k-2}{2}}D^{\frac{k}{2}}\ + \\
..
\end{array} \notag
\end{align}
When even \(k\to\infty\),
\begin{align}
\psi\left(\mathbf B_{d},p\right)&=\lim_{k\to\infty}n_{k}\Big(\mathcal A_{k}\left(\mathbf B_{d}\right)\Big)\cdot p^{k} \notag
\end{align}
\begin{align}
\psi\!\left(\mathbf B_{d},p\right)&=\!\!\lim_{k\to\infty}\!\left\{\!\!
\begin{array}{c}
\displaystyle\Bigg(\left(\frac{D}{2}\right)^{\frac{1}{2(k-1)}}\left(2D\right)^{\frac{1}{2}}p\Bigg)^{k-1}+ \\
\displaystyle\Bigg(\!\!\left(\frac{1}{2D}\right)^{\frac{1}{2(k-1)}}{\frac{k}{2}\choose1}^{\frac{1}{k-1}}\left(2D\right)^{\frac{1}{2}}p\Bigg)^{k-1}+ \\
\displaystyle\Bigg(\!\!\left(\!\frac{1}{2^{3}D}\!\right)^{\frac{1}{2(k-1)}}{\frac{k}{2}\choose2}^{\frac{1}{k-1}}\left(2D\right)^{\frac{1}{2}}p\!\Bigg)^{k-1}\!\!+ \\
\displaystyle\Bigg(\!\!\left(\!\frac{1}{2^{3}D^{3}}\!\right)^{\frac{1}{2(k-1)}}\!\!{\frac{k+2}{2}\choose3}^{\frac{1}{k-1}}\!\!\left(2D\right)^{\frac{1}{2}}p\!\Bigg)^{k-1}\!\!+ \\ \\
..
\end{array}
\right. \notag
\end{align}
or
\begin{align}
\psi\left(\mathbf B_{d},p\right)&=\lim_{k\to\infty}\left\{
\begin{array}{c}
\Big(\left(2D\right)^{\frac{1}{2}}p\Big)^{k}\ + \\
\displaystyle\Bigg(\left(\frac{1}{2}\right)^{\frac{1}{k}}{\frac{k}{2}\choose1}^{\frac{1}{k}}\left(2D\right)^{\frac{1}{2}}p\Bigg)^{k}\ + \\
\displaystyle\Bigg(\left(\frac{1}{2D}\right)^{\frac{1}{k}}{\frac{k+2}{2}\choose2}^{\frac{1}{k}}\left(2D\right)^{\frac{1}{2}}p\Bigg)^{k}\ + \\
\displaystyle\Bigg(\left(\frac{1}{2^{2}D}\right)^{\frac{1}{k}}{\frac{k+2}{2}\choose3}^{\frac{1}{k}}\left(2D\right)^{\frac{1}{2}}p\Bigg)^{k}\ + \\ \\
..
\end{array}
\right. \notag
\end{align}
For \(D=2d-2\) and \(d\ge2\), \(p_{H}\left(\mathbf B_{d}\right)\) is the smallest \(p\) for which \(\psi\left(\mathbf B_{d},p\right)\ge1\):
\begin{align}
p_{H}\left(\mathbf B_{d}\right)&=\frac{1}{2(d-1)^{\frac{1}{2}}}\cdot\text{min}\left\{
\begin{array}{c}
\displaystyle\lim_{k\to\infty}\left((d-1)^{-\frac{1}{2}}\right)^{\frac{1}{k-1}} \\ \\
\ \ \ \ \ \text{or} \\ \\
\displaystyle\lim_{k\to\infty}\left(\frac{2(d-1)^{\frac{1}{2}}}{\displaystyle{\frac{k}{2}\choose1}}\right)^{\frac{1}{k-1}} \\ \\
\ \ \ \ \ \text{or} \\ \\
\displaystyle\lim_{k\to\infty}\left(\frac{2^{2}(d-1)^{\frac{1}{2}}}{\displaystyle{\frac{k+2}{2}\choose2}}\right)^{\frac{1}{k-1}} \\ \\
\ \ \ \ \ \text{or} \\ \\
\displaystyle\lim_{k\to\infty}\left(\frac{2^{3}(d-1)^{\frac{3}{2}}}{\displaystyle{\frac{k+2}{2}\choose3}}\right)^{\frac{1}{k-1}} \\ \\
\ \ \ \ \ \text{or} \\ \\
..
\end{array}
\right. \notag
\end{align}
or
\begin{align}
p_{H}\left(\mathbf B_{d}\right)&=\frac{1}{2(d-1)^{\frac{1}{2}}}\cdot\text{min}\left\{
\begin{array}{c}
1 \\ \\
\ \ \ \ \ \text{or} \\ \\
\displaystyle\lim_{k\to\infty}\left(\frac{2}{\displaystyle{\frac{k}{2}\choose1}}\right)^{\frac{1}{k}} \\ \\
\ \ \ \ \ \text{or} \\ \\
\displaystyle\lim_{k\to\infty}\left(\frac{2^{2}(d-1)}{\displaystyle{\frac{k+2}{2}\choose2}}\right)^{\frac{1}{k}} \\ \\
\ \ \ \ \ \text{or} \\ \\
\displaystyle\lim_{k\to\infty}\left(\frac{2^{3}(d-1)}{\displaystyle{\frac{k+2}{2}\choose3}}\right)^{\frac{1}{k}} \\ \\
\ \ \ \ \ \text{or} \\ \\
..
\end{array}
\right. \notag
\end{align}
For \(0\le m\le n\) and \(m,n\le k\) and \(k\to\infty\),
\begin{align}
\notag \\
&\text{min}\left\{\displaystyle\lim_{k\to\infty}\left(\frac{2^{n}(d-1)^{m}}{\displaystyle{\frac{k+n}{2}\choose n}}\right)^{\frac{1}{k}}\right\}=1 \notag \\ \notag \\
&\text{min}\left\{\displaystyle\lim_{k\to\infty}\left(\frac{2^{n}(d-1)^{\frac{m}{2}}}{\displaystyle{\frac{k+n}{2}\choose n}}\right)^{\frac{1}{k-1}}\right\}=1 \notag
\end{align}
\(p_{H}\left(\mathbf B_{1}\right)=1\) and for \(2\le d\le k\) and \(k\to\infty\), 
\begin{align}
&\boxed{p_{H}\left(\mathbf B_{d}\right)=\displaystyle\frac{1}{2(d-1)^{\frac{1}{2}}}} \notag
\end{align}

\bibliography{bibliography}
\bibliographystyle{unsrt}
\end{document}